\documentclass[12pt,a4paper]{article}
\usepackage{amsmath,amssymb}
\usepackage{setspace}
\usepackage{axodraw}
\usepackage{slashed}
\usepackage{cite,hyperref}
\usepackage{mathtools}
\usepackage[margin=1in]{geometry}
\usepackage{titlesec}
\usepackage{xcolor}

\titleformat{\section}
  {\normalfont\Large\bfseries}{\thesection}{1em}{}[{\color{gray}\titlerule[0.4pt]}]

\def\la{\langle}
\def\ra{\rangle}
\def\spa#1.#2{\left\langle#1\,#2\right\rangle}
\def\spb#1.#2{\left[#1\,#2\right]}
\def\spba#1.#2.#3{\left[#1|#2|#3\right\rangle}

\newcommand{\Yy}{\Upsilon}
\newcommand{\Ss}{S}

\newcommand{\pr}{Q}

\newcommand{\mi}{\scalebox{0.94}[1.0]{-}}

\DeclareMathOperator{\tr}{\mathrm{tr}}

\def\notag{\nonumber}

\usepackage{authblk}
\author[1]{Guy~R.~Jehu}
\affil[1]{\textit{School of Mathematics and Hamilton Mathematical 
Institute,}}
\affil[ ]{\textit{Trinity College Dublin, College Green,
Dublin, Ireland}}
\affil[ ]{{\tt \color{blue}\href{mailto:jehu@maths.tcd.ie}{jehu@maths.tcd.ie}}}
\title{Symmetric reduction of high-multiplicity one-loop integrals and maximal cuts}
\date{\today}
\numberwithin{equation}{section}
\begin{document}
\maketitle
\begin{abstract}
	We derive useful reduction formulae
	which express one-loop 
	Feynman integrals
	with a large number of external momenta in terms of 
	lower-point integrals carrying easily derivable
	kinematic coefficients which are
	symmetric in the external momenta.
	These formulae apply for integrals
	with at least two more external legs 
	than the dimension of the external momenta,
	and are presented in terms of  
	two possible bases: 
	one composed of a subset of descendant integrals 
	with one fewer external legs,
	the other composed of
	the complete set of minimally-descendant
	integrals with just one more leg than the dimension of
	external momenta.
	In 3+1 dimensions, particularly compact 
	representations of kinematic
	invariants can be computed, which easily lend
	themselves to spinor-helicity or trace representations.
	The reduction formulae have a close relationship with 
	$D$-dimensional unitarity cuts, 
	and thus provide a path towards computing
	full (all-$\epsilon$) expressions for scattering amplitudes
	at arbitrary multiplicity.
\end{abstract}
\newpage
\section{Introduction}
The computation of multi-loop helicity amplitudes in Yang-Mills theories has recently
met with great success, such as at two loops, five points for 
QCD~\cite{Badger:2015lda,Badger2018,Abreu:2019odu,Badger:2019djh,Dunbar2020a}, and, 
using myriad mathematical tools, through to seven loops and six points for planar $\mathcal{N}=4$ 
super Yang-Mills~\cite{Caron-Huot:2019bsq}.
Despite this, all multiplicity (all-$n$) analytic forms of one-loop amplitudes in Yang-Mills theory are known only
up to $\mathcal{O}(\epsilon)$, and only in cases with maximal supersymmetry and MHV  helicity structure~\cite{Bern:1994cg},
or all-plus and single-minus helicity configurations~\cite{Mahlon:1993si,Mahlon:1994dc}.
Recently, an all-$n$ form for a two-loop partial all-plus amplitude in pure Yang-Mills 
has been computed~\cite{Dunbar2020}.
One of the principle barriers to computing these is a lack of 
a concise analytic understanding of one-loop 
high-multiplicity integrals.

One way this is relevant to contemporary multi-loop computations is
that all-$\epsilon$ analytic expressions 
occur when applying one-loop 
results to multi-loop integrals~\cite{Anastasiou2003,Bern2005}; 
recent computations of all-plus amplitudes at two loops reveal a tantalisingly one-loop like
analytic structure~\cite{Gehrmann:2015bfy,Dunbar:2016aux,Dunbar:2017nfy,Badger:2019djh,Henn:2019mvc}.
Moreover, higher-in-epsilon terms of one-loop amplitudes
contribute at two-loops
Such an amplitude $A^{{\rm 2-loop}}_n$
can be arranged in a way that 
manifests its universal infra-red (IR) 
structure~\cite{Catani:1998bh}:
\begin{align}
	A_n^{{\rm 2-loop}} = A_n^{{\rm tree}} \mathfrak{I}^{2}
			+A_n^{{\rm 1-loop}}\mathfrak{I}^{1}
			+F_n,
\end{align}
where the universal infra-red terms $\mathfrak{I}^{m}$ are
defined in~\cite{Catani:1998bh}.
In particular, the term $\mathfrak{I}^{1}$
contains infra-red poles in the dimensional regulator
up to $\mathcal{O}(\epsilon^{-2}) $ which would cancel with
the higher-in-$\epsilon$ terms in the one-loop amplitude 
$A^{{\rm 1-loop}}$ upon expansion,
leaving sub-leading and extra finite terms in 
addition to $F_n$.

Beyond this, the all-$\epsilon$ one-loop structure 
of $\mathcal{N}=4$ super Yang-Mills
is conjectured to be proportional to a dimension-shifted all-plus amplitude in pure Yang-Mills~\cite{Bern:1996ja};
 a conclusive proof of this is still outstanding, but will be 
 addressed in a forthcoming paper~\cite{Britto:2020crg}.
Generalising results such as these which rely on 
particular contingencies requires
a more complete understanding of the all-$n$
analytic structure of both tree amplitudes and integrals; 
this paper focuses on one-loop instances of the latter.

Techniques to reduce finite $n$-point one-loop Feynman integrals 
(from here on referred to as an $n$-gon) in 
integer dimension have long been
well understood~\cite{Brown1961,Halpern1963,Melrose1965,
Petersson1965,Neerven1984}. 
The use of dimensional regularisation 
to deal with both ultraviolet (UV) and
infra-red (IR) divergencies led to  
more general formulae~\cite{Bern:1992em,Bern:1993kr} 
which both justify a small basis of 
scalar integrals for one-loop amplitudes 
and codify relations amongst the integrals as a particular case
of dimensional recurrence identities~\cite{Tarasov1996d}.
These
can in turn be used to solve the integrals 
themselves~\cite{Tarasov1996d,Lee2010a,Lee2010b,Lee2012,Bytev2014}.

These formulae face issues, however, when $n>2+L$, for  $L$ 
the dimension of the 
external momenta\footnote{We assume that the external legs are
always kept in integer dimension, for example in 
the four-dimensional
helicity scheme of dimensional regularisation~\cite{Bern:1991aq}} 
, in which case 
the derivation depends on the inversion of a singular matrix $S$.
A
formalism was in turn developed~\cite{Binoth2000} in terms of defining a pseudo-inverse of $S$
to resolve high-$n$ integrals into lower-point ones.
Although suitable for application in the form of computer algorithms,
the approach breaks the cyclic symmetry of the external legs,
obscuring the simplicity of the analytic structure 
and making it difficult to apply to possible conjectures of all-$n$ all-$\epsilon$ forms.

In this paper we present an alternative which preserves the symmetries when reduced to 
$(L+1)$-gons, and allows simple basis choices to be made for the reduction to $(n-1)$-gons.
For the equal propagator-mass case where $L=4$, 
the reduction formula from $n$-point 
integrals\footnote{Normalisations and
definitions are found in section~\ref{sec:rev}.}
$I_n$ to 
pentagon integrals $I_5$ is
especially simple and takes the form
\begin{align}
	I_n &= {1\over 2^{n-5}}\sum_{i_1,...,i_{n-5}=1}^{n}\left[\prod_{m}^{n-5}
	\xi_{i_m}^{\left[\mathcal{P}_i, i_m\right]}\right]
	I_{5}^{\left[\mathcal{P}_i \right]} \; ,
	\notag \\
	\mathcal{P}_{i} &= \lbrace 1,...,n\rbrace\slash 
	\left\lbrace i_1,...,i_{n-5}\right\rbrace\; ,
	\label{eq:pentred}
\end{align}
where
\begin{equation}
	\xi^{[k_1,k_2,k_3,k_4,k_5,k_6]}_{k_j} = (-1)^{j}{2\tr_5 \left(q_{k_{j+1}k_{j+2}} 
	q_{k_{j+2}k_{j+3}}q_{k_{j+3}k_{j+4}}q_{k_{j+4}k_{j+5}}\right) 
	\over 
	\tr_5 \left(q_{k_1k_2} q_{k_2k_3}q_{k_3k_4}q_{k_4k_5}q_{k_5k_6}q_{k_6k_1}\right) } \; ;
\end{equation}
here $\tr_5(p_ap_b... ) = \tr(\gamma_5 \slashed{p}_a\slashed{p}_b... ) $, $q_{ij}= p_i+p_{i+1}+...+p_{j-1}$ where $p_i$ is the (cyclically indexed) $i$th outgoing external momentum.

The paper is structured as follows: In section~\ref{sec:rev}
the notation is introduced, and we review derivations of some
previously known results.
In section~\ref{sec:dfour} we explicitly study the 
the case where the 
external dimensions are in four-dimensional Minkowski space,
giving explicit analytic expressions for the various 
kinematic determinants and coefficients for the 
pentagon and hexagon cases.
In 
section~\ref{sec:newform} we derive new formulae 
for reducing loop integrals
in the singular cases where kinematic determinants vanish.
In section~\ref{sec:cuts}, the connection to a type of maximal 
cut on the pentagon is discussed.

\newpage
\section{Review and definitions}
\label{sec:rev}
\begin{figure}[ht]
\centerline{
    \begin{picture}(300,190)(-150,-60)    
    \Line(-60,60)(0,90)
	    \LongArrow(-50,57)(-4,80)
	    \Text(-20,56)[r]{$\ell$}
    \Line(0,90)(60,60)
    \Line(60,60)(90,10)
	    \DashLine(90,10)(20,-40){3}
	    \DashLine(-60,60)(-60,0){3}
	    \SetWidth{1.5}
	    \Line(-60,60)(-70,70)
	    \Line(0,90)(0,102)
	    \Line(60,60)(70,70)
	    \Line(90,10)(102,10)
	    \SetWidth{1}
	    \Text(-73,70)[r]{$p_n$}
	    \Text(0,110)[c]{$p_1$}
	    \Text(74,74)[l]{$p_2$}
	    \Text(105,12)[l]{$p_3$}
    \end{picture} 
    }
	\caption[Generic Feynman integrals]{Diagrammatic representation of the general 
	one-loop integral function $\mathcal{I}^D_n$ in eq.~\ref{eq:masint}. The external momenta $p_i$ should
	always be considered outgoing.}
    \label{fig:masint}
\end{figure}
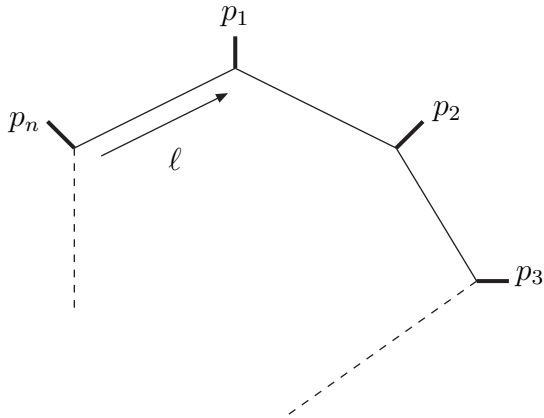

Basing our notation and analysis on~\cite{Bern:1992em,Bern:1993kr},
a general $D$-dimensional one-loop scalar integral corresponding to the momentum routing in figure~\ref{fig:masint} is defined as
\begin{align}
	\mathcal{I}^D_n &=
	\int{d^D\ell\over(2\pi)^D}
	{1\over (\ell^2-M_1^2)((\ell-q_2)^2-M_2^2)\cdots ((\ell-q_{n})^2-M_{n}^2)} \; ,
    \label{eq:masint}
\end{align}
where the momenta 
\begin{equation}
	q_i= p_1+p_2+...+p_{i\mi 1} \quad .
\end{equation}

Upon integration, the denominator can be expressed as a homogeneous 
second-order polynomial of Feynman parameters $a_i$
\begin{align}
	i(\mi 1)^{n+1}(4\pi)^{D/2}\mathcal{I}^D_n &= \Gamma(n-D/2)
	\int_0^1d^na_i\delta(1-\Sigma_ia_i)
	{1\over \bigl[\sum_{i,j=1}^n a_i\Ss_{ij}a_j\bigr]^{n-D/2}}
	\notag \\
	&\equiv I_n[1]\; ;
\end{align}
where the $n$-gon matrix\footnote{In terms of graph polynomials,
this is just a representation the $F$ (second Symanzik) 
polynomial as a bilinear map of the Feynman parameters.}
\begin{equation}
	\Ss_{ij}={1\over 2} \left( M_i^2+M_j^2-q_{ij}^2\right)\; ,
\end{equation}
with $q_{ij}$ defined as
\begin{align}
	q_{ij} = q_{j}-q_{i} = \sum_i^{j\mi 1} p_i \; ;
\end{align}
these fully characterise the integral together with the 
dimension $D$ and the numerator which we express in the argument 
of $I_n$ and is in general a monomial of Feynman parameters.

To avoid repeated explicit denotations of determinants, 
we introduce
\begin{equation}
\Yy_n\equiv\det \Ss \; ,
\end{equation}
and the Gram determinant,
	$\Delta_n \equiv \det[2p_i\cdot p_j] $,
can be expressed in terms of the $n$-gon matrix as a Cayley-Menger
determinant:
\begin{equation}
	\Delta_n = 2^{n-1}\left| 
	\begin{array}{cc}
		0 & -{\bf 1}^T \\
		{\bf 1} & \Ss
	\end{array}
	\right|\; ,
	\label{eq:caymen}
\end{equation}
where {\bf 1} here represents a column of $n$ entries of value $1$.

We can derive linear relations between integrals that are useful
in practical calculations by considering alternative ways of deriving the same 
quantities~\cite{Bern:1993kr}.
Consider the cancellation of a propagator in eq.~\ref{eq:masint}:
\begin{align}
	J_1 =i(-1)^{n+1}(4\pi)^{D/2} 
	\int{d^D\ell\over(2\pi)^D}
	{\ell^2-M_1^2\over (\ell^2-M_1^2)((\ell-q_1)^2-M_2^2)\cdots ((\ell-q_{n-1})^2-M_{n}^2)} \; ;
\end{align}
whence we can derive a relation by 
considering the evaluation of this integral in two 
different ways.
Firstly, the obvious
\begin{align}
	J_1 = -I^{(1)}_{n-1}[1]
\end{align}
comes from a cancellation with the propagator;
the superscript $(1)$ here means pinching the propagator between $n$ and $1$.
Secondly, leaving the numerator in place and Feynman parametrising leads to the identity
\begin{align}
	-I^{(1)}_{n-1}[1]= \left(n-1-D\right)
		I^{D+2}_n[1]- 2\sum_{j=1}^n\Ss_{1j}I_n[a_j]\; .
		\label{eq:fo2}
\end{align}
Generalising to any inverse propagator in the numerator
and, where
$\Upsilon_n \neq 0$, inverting gives
\begin{align}
	I_{n}[a_i]= {1\over 2}\left[\sum_{j=1}^n\Ss^{\mi 1}_{ij}I^{(j)}_{n-1}[1]
		+(n-1-D)\sum_{j=1}^n\Ss^{\mi 1}_{ij}I^{D+2}_n[1]\right] .
		\label{eq:onepam}
\end{align}
Summing the left-hand side using $\sum_{i=1}^n a_i = 1$
gives
\begin{align}
	I_n[1]= {1\over 2}\left[\sum_{j=1}^nc_jI^{(j)}_{n-1}[1]
		+(n-1-D)c_0I^{D+2}_n[1]\right]\;  ,
		\label{eq:red}
\end{align}
where we have now defined
\begin{equation}
	c_i = \sum_{j=1}^n\Ss^{\mi 1}_{ij} 
\end{equation}
and
\begin{align}
	c_0&\equiv\sum_{j=1}^n c_j 
	\notag \\
	&= (2)^{1- n} {\Delta_n\over \Upsilon_n}\quad .
	\label{eq:cno}
\end{align}

Equation~\ref{eq:red} clearly diverges if $\Upsilon_n = 0$.
Introducing $L$ for the dimension of the external momenta,
then the condition on the vanishing of $\Upsilon_n$ is
\begin{equation}
	  n>L+2 \quad .
	\label{eq:cond}
\end{equation}
We explore the case where $L=4$ in section~\ref{sec:dfour}, 
and derive the formula for the general-$L$ cases defined
by equation~\ref{eq:cond} 
in section~\ref{sec:newform}.

Assuming generic 
parameters $\lbrace M_i,q_i\rbrace$ to avoid IR divergences
allows us to set $D$ to be an integer, and
leads to the integer version of the $n=L+1$ reduction formula
\begin{equation}
	I_{L+1}[1]= {1\over 2}\sum_{j=1}^nc_jI^{(j)}_{L}[1]\quad .
\end{equation}
This case was proved long 
ago~\cite{Brown1961,Halpern1963,Melrose1965,
Petersson1965,Neerven1984}, and indeed
the generalisation to higher $n$ was 
also proved, first by Melrose~\cite{Melrose1965}
and very shortly afterwards by Petersson~\cite{Petersson1965}.
Continuation of this to the case where $D$ is not an integer is
the main result of this paper.

It is worthwhile highlighting
that Petersson~\cite{Petersson1965}
defines the $c_i$ in a different manner:
\begin{equation}
	c_i = {1\over (l_i^+-q_{i-1})^2-M_i}+ 
	{1\over (l_i^+-q_{i-1})^2-M_i}
	\; ,
	\label{eq:peter}
\end{equation}
where $l_i^\pm $ are the two solutions to the $L$ equations
\begin{align}
	(\ell-q_{j-1})^2 -M_j^2 =  0  \quad j\neq i \quad .
\end{align}
These can be recognised as maximal 
cut constraints~\cite{Britto:2004nc}
for the $L$-point integral $I_{L}^{(i)}$,
although they were not considered as such at the time.
This is expanded upon for the $L=4$ case 
and more generally to non-integer $D$
in section~\ref{sec:cuts}.

\newpage
\section{Four-dimensional Minkowski space}
The various kinematic coefficients ($c_i$, $\Upsilon_n$, 
$\Delta_n$ etc...) can of course be enumerated using computer
algebra systems, however we show in this section that
it is in fact quite easy to derive conveniently compact 
expressions
where $L=4$ and the internal masses are degenerate.
\label{sec:dfour}
\subsection{Pentagon}
Our first example  is the version of 
the shift relation~\ref{eq:red} 
which reduces the pentagon, $I_5$,
to boxes, $I^{(j)}_4$, and
a dimension shifted pentagon
\begin{equation}
	I_5= {1\over 2}\left[\sum_{j=1}^5c_jI^{(j)}_{4}
		+(4-D)c_0I^{D+2}_5\right] \quad .
\end{equation}
The Gram determinant is given by 
\begin{align}
	\Delta_5 &=\det_{i,j\neq 5}[2p_i\cdot p_j]  
	\notag \\
	&= 16\det_{i\neq 5}[p_{i\mu}] \det_{j\neq 5}[ p^\mu_{j}] 
	\notag \\
	&=-16\varepsilon^{p_1p_2p_3p_4}	\varepsilon_{p_1p_2p_3p_4}
	\notag \\
	&=\tr^2_5(1234)\equiv \left(\tr(\gamma_5\slashed{p}_1\slashed{p}_2\slashed{p}_3\slashed{p}_4)\right)^2\quad .
\end{align}
For simplicity considering the all-massless (i.e. propagators as well as external momenta)
case, then the pentagon determinant is, 
written in Mandlestam ($s_{ij}=(p_i+p_j)^2$) notation,
\begin{equation}
	\Upsilon_5 =\left(-{1\over 2}\right)^{5}
	\left|
	\begin{array}{ccccc}
		0 & 0 & s_{12} & s_{45} & 0 \\
		0 & 0 & 0 & s_{23} & s_{51} \\
		s_{12} & 0 & 0 & 0 & s_{34} \\
		s_{45} & s_{23}  & 0 & 0 & 0 \\
		0 & s_{51}  & s_{34} & 0 & 0 
\end{array}
        \right|
		=-{1\over 16} s_{12}s_{23}s_{34}s_{45}s_{51} \quad .
\end{equation}
The $c_i$'s can readily be determined by noting that, from conservation of momentum
\begin{align}
	&-{1\over 2}\left(
	\begin{array}{ccccc}
	0&0&1&1&0 \\
	0&0&0&1&1 \\
	1&0&0&0&1 \\
	1&1&0&0&0 \\
	0&1&1&0&0 \\
	\end{array}
	\right)
	\left(
	\begin{array}{c}
		\tr(4512) \\
		\tr(5123) \\
		\tr(1234) \\
		\tr(2345) \\
		\tr(3451) 
	\end{array}\right)
	 = 
	\left( \begin{array}{c}
		s_{23}s_{34} \\
		s_{34}s_{45} \\
		s_{45}s_{51} \\
		s_{51}s_{12} \\
		s_{12}s_{23} 
\end{array}
\right)
\end{align}
and as 
\begin{equation}
	\Ss_{ij} c_j = 1 \; ,
\end{equation}
we can identify
\begin{equation}
	c_i = -{1\over 16\Upsilon_5} s_{(i+1)(i+2)}s_{(i+2)(i+3)} \tr\big((i-2)(i-1)i(i+1)\big) \quad .
	\label{eq:cpent}
\end{equation}
The indices take on a cyclic definition (i.e. $i+1=1$ where $i=5$). 

\subsection{Hexagon}
At $n=6$ the reduction formula~\ref{eq:red} is 
simplified by the fact that $\Delta_6 = 0$ for $L=4$,
and thus a hexagon can be written entirely in terms of 
pentagons~\cite{Bern:1993kr}
\begin{align}
	I_6[1]= {1\over 2}\sum_{j=1}^6c_jI^{(j)}_{5}[1] \; .
	\label{eq:hex}
\end{align}
It is useful to have a compact value for $\Upsilon_6$. 
It was shown explicitly~\cite{Binoth2005} that for the all-massless case
\begin{equation}
	\Upsilon_6 = -{1\over 64}\tr_5^2(123456) \; .
\end{equation}
This is in fact also the case for the generic
external mass ($\lbrace p_i^2=m_i^2\rbrace$) and equal internal mass ($\lbrace M_i^2= M^2\rbrace$) case;
this can be shown by considering
the Cayley-Menger matrix $C$
\begin{align}
	C = \left( 
	\begin{array} {cc}
		0 & -{\bf 1} \\
		\bf{1} & \Ss
	\end{array}
	\right) \quad .
\end{align}
The minors of $C$ can be identified for the diagonal elements,
by using the fact that they correspond pentagon Cayley-Menger 
determinants which, from equation~\ref{eq:caymen},
are proportional to Gram determinants.
Explicitly the adjugate to the matrix $C$ can be partially filled
in
\begin{align}
	C^{{\rm adj}}  
	&=
	 \left( 
	\begin{array}{ccccccc}
		\Upsilon_6&*&*&*&*&*&*               \\
		*&\mi 32\Delta_5^{(1)}&*&*&*&*&* \\
		*&*&\mi 32\Delta_5^{(2)}&*&*&*&* \\
		*&*&*&\mi 32\Delta_5^{(3)}&*&*&* \\
		*&*&*&*&\mi 32\Delta_5^{(4)}&*&* \\
		*&*&*&*&*&\mi 32\Delta_5^{(5)}&* \\
		*&*&*&*&*&*&\mi 32\Delta_5^{(6)} \\
	\end{array}
	\right) \; ,
\end{align}
where $\Delta_5^{(j)}$ is the Gram determinant of the pentagon with the
$j$th propagator pinched, and $*$ represents entries yet to be determined.

From the fact that $C$ is a symmetric 7x7 matrix of rank $6$,
$C^{{\rm adj}}$ must be a symmetric rank 1 matrix,
thus expressible as the outer product of two instances 
of the same vector $ b= (b_0, {\bf b})^T$ 
\begin{equation}
	C_{ij} = { b}_i{b}_j  \; i,j \in\lbrace 0,...,6\rbrace\; 
	,
\end{equation}
where $ b $ satisfies the conditions
\begin{align}
	b_1^2 &= \Upsilon_6 \; ,
	\label{eq:con1}
	\\
	b_i^2 &= 32\Delta_5^{(i-1)} \quad i\neq 0 \quad .
	\label{eq:con2}
\end{align}
As 
\begin{equation}
	CC^{{\rm adj}}={\rm Diag} (\Delta_6) = 0 \; ,
\end{equation}
we can deduce from equation~\ref{eq:cno} that
\begin{align}
	{\bf 1} \cdot {\bf b} &= \sum_{j=1}^6b_j= 0  \; ,
	\label{eq:con3}
	 \\
	\sum_{j=1}^6 \Ss_{ij}\cdot   b_{j+1} &=  -{b_0} \quad.
\label{eq:con4}
\end{align}
The latter two conditions imply that 
\begin{equation}
	c_i = -{b_{i}\over b_0} \; .
	\label{eq:match}
\end{equation}

For equal mass propagators the identity 
\begin{align}
	\sum_{j=1}^6\Ss_{ij}(-1)^j\tr_5\big((j+1)(j+2)(j+3)(j+4)\big) = {1\over 2} \tr_5(123456)
\end{align}
can be used to deduce
that
\begin{align}
	c_i& = (-1)^i{2\tr_5\big( (i+1)(i+2)(i+3)(i+4)\big)\over \tr_5(123456)}  
	\notag \\
	\Rightarrow
	c_i^2& =4 {\Delta_5^{(i)}\over \tr_5^2(123456)}
\end{align}
and matching with equations~\ref{eq:match}, \ref{eq:con1} and~\ref{eq:con2} gives
\begin{equation}
	\Upsilon_6 = -{1\over 64} \tr^2_5(123456) \quad .
\end{equation}
This remarkably compact 
form of the hexagon determinant is valid for all cases where
the external momenta $p_i$ are massive or null, 
but the propagators have equal mass.
These expressions are practically useful for 
computing loop amplitudes or form factors with six or more 
external legs and equal internal masses, especially when paired
with the analysis in the following sections.
A Petersson style analysis in terms of 
the unparametrised Feynman integral is 
carried out in section~\ref{sec:cuts}.

\newpage
\section{Reduction formulae}
\label{sec:newform}
We begin this section by introducing a specific notation for
the kinematic coefficients of the reduction formula~\ref{eq:red}
in the
$L+2\rightarrow L+1$ case
\begin{align}
	\hspace{2in}\xi_i\equiv c_i \quad {\rm for\;}L+2\rightarrow L+1-{\rm point\;reduction}
	\quad .
\end{align}
 In this section we show that these particular variables are all
that are needed for the $n\geq L+2$ reduction.
In particular we show that the coefficients of reduction
consist simply of a product of the $\xi$s corresponding to
all possible descendant  
$(L+2)$-gons of the given $n$-gon.
Conversely, a reduction to $(n-1)$-gons requires 
a choice of $(L+2)$-dimensional basis 
with coefficients corresponding to the $\xi$s from a reduction
of the  $(L+2)$-gon complementary to said basis;
this is illustrated in figure~\ref{fig:hex} below.
We also explain how to deal with Feynman
parameters in the numerator.

\subsection{Generalised reduction of scalar integrals}
The $L$-dimensional version of equation~\ref{eq:hex} is 
\begin{equation}
	I_{L+2} = {1\over 2}\sum_{j=1}^{L+2} \xi_j I^{(j)}_{L+1} \; , 
	\label{eq:lred}
\end{equation}
and with this notation, the reduction of a general descendant
$L+2$-gon is
\begin{equation}
	I^{(i_1,i_2,...,i_{n-L-2})}_{L+2} 
	= \sum_{j\in \mathcal{E}^{\left(i_1,i_2,...i_{n-L-2}\right)}_{L+2}} 
	\xi^{(i_1,i_2,...,i_{n-L-2})}_j I^{(i_1,i_2,...,i_{n-L-2},j)}_{L+1} \; ;
\end{equation}
where $\mathcal{E}^{(i_1,i_2,...,i_{n-L-2})}_{L+2}$ is the 
set
$\lbrace 1,...,n\rbrace \slash \lbrace i_1,i_2,...i_{n-L-2}\rbrace$
which indexes the descendant $(L+2)$-gon.

It should be emphasised that there are 
two notation schemes for integral
reduction:
\textbf{descendant notation} indexed by the propagators shrunk
from the ancestor $n$-gon, and 
the \textbf{ascendant notation} indexed by the propagators 
expanded.
To translate between the two
\begin{align}
	I^{(i_1,i_2,...,i_{n-L-2})}\equiv 
	I^{[\bar{i}_1,\bar{i}_2,...,\bar{i}_{L+2}]}
	\quad {\rm for} \; \bar{i}_k\in
	\mathcal{E}^{(i_1,i_2,...,i_{n-L-2})}_{L+2}\; ,
\end{align}
where the descendant notation is on the left and 
the ascendant notation is on the right.

To generalise equation~\ref{eq:lred} to $n>L+2$, 
we start by looking at the $n=L+3$ case. Equation~\ref{eq:fo2}
gives
\begin{align}
	-I_{L+2}^{(i)}[1]&= \left(L+1-D\right)
		I^{D+2}_{L+3}[1]
		- 2\sum_{j=1}^n\Ss_{ij}I_{L+3}[a_j]\; ,
		\label{eq:startder}
\end{align}
which with some linear algebra (shown in appendix~\ref{app:der}) 
leads to
\begin{align}
	 I_{L+3}={1\over 2}
	 \sum_{j\in\mathcal{E}_{L+2}}\xi^{(k)}_iI_{L+2}^{(i)} \; .
	 \label{eq:endder}
\end{align}
The manifest symmetry is broken by an explicit choice of descendant
$(L+2)$-gon (the choice of pinched propagator $k$), 
however it is restored at the level of $(L+1)$-gons
\begin{align}
	I_{L+3}={1\over 4}\sum_{j,k=1}^n\xi^{(j)}_k\xi^{(k)}_jI_{L+1}^{(j,k)} \; .
	\label{eq:hept}
\end{align}

\begin{figure}[ht]
\centerline{
    \begin{picture}(240,150)(-120,-70)    
	    \Line(20,-30)(80,-30)
	    \Line(20,-30)(-10,22)
	    \Line(80,-30)(110,22)
	    \Line(20,74)(80,74)
	    \Line(20,74)(-10,22)
	    \Line(80,74)(110,22)
	    \Line(20,-30)(20,-60)
     		\Text(20,-66)[r]{$k_1$}   
	    \Line(20,-30)(-7,-42)
	    \Line(80,-60)(80,-30)
	    \Line(80,-30)(107,-42)
	    \Line(20,74)(20,104)
	    \Line(80,74)(107,86)
	    \Line(20,74)(-7,86)
     		\Text(-20,88)[l]{$k_3$}   
     		\Text(80,109)[l]{$k_4$}   
     		\Text(133,38)[l]{$k_5$}   
     		\Text(110,-42)[l]{$k_6$}   
	    \Line(80,74)(80,104)
	    \Line(-10,22)(-30,38)
     		\Text(-32,9)[r]{$k_2$}   
	    \Line(110,22)(130,38)
	    \Line(-10,22)(-30,6)
	    \Line(110,22)(130,6)
   \CCirc(5,-45){1.5}{Black}{Black} 
   \CCirc(13,-51){1.5}{Black}{Black} 
   \CCirc(95,-45){1.5}{Black}{Black} 
   \CCirc(87,-51){1.5}{Black}{Black} 
   \CCirc(5,89){1.5}{Black}{Black} 
   \CCirc(13,95){1.5}{Black}{Black} 
   \CCirc(95,89){1.5}{Black}{Black} 
   \CCirc(87,95){1.5}{Black}{Black} 
   \CCirc(120,25){1.5}{Black}{Black} 
   \CCirc(120,19){1.5}{Black}{Black} 
   \CCirc(-20,25){1.5}{Black}{Black} 
   \CCirc(-20,19){1.5}{Black}{Black} 
    \end{picture} 
    \begin{picture}(300,190)(-130,-60)    
    \Line(-60,60)(0,90)
    \Line(0,90)(60,60)
    \Line(60,60)(90,10)
	    \DashLine(90,10)(20,-40){3}
	    \DashLine(-60,60)(-60,0){3}
	    \SetWidth{1}
	    \Line(-60,60)(-70,70)
	    \Line(0,90)(0,102)
	    \Line(60,60)(78,88)
	    \Line(60,60)(90,77)
	    \Line(90,10)(102,10)
	    \SetWidth{1}
	    \Text(-73,70)[r]{$k_i-3$}
	    \Text(0,110)[c]{$k_i-2$}
	    \Text(82,94)[l]{$k_i-1$}
	    \Text(96,74)[l]{$k_i$}
	    \Text(105,12)[l]{$k_i+1$}
    \end{picture} 
    }
	\caption[Hexagon]{LEFT: A choice of hexagon is determined by a choice of six external
	legs of $I_n$. 
	The coefficients are the $\xi_i$s corresponding to the reduction of this hexagon to pentagons.

	RIGHT: The basis of the reduction it corresponds to are the six ($n-1$)-gons which are not ancestors of this hexagon. 
	  }
    \label{fig:hex}
\end{figure}
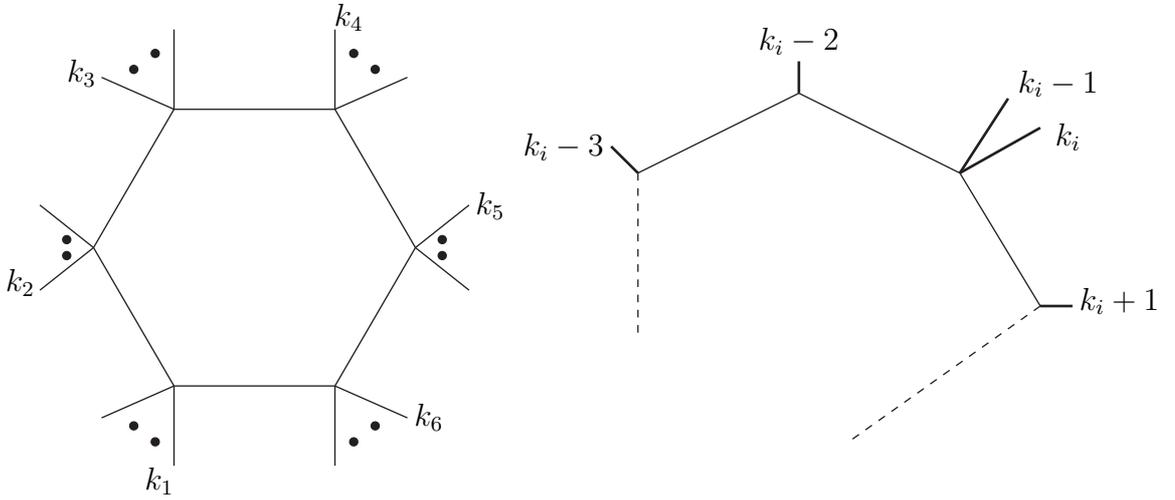

The derivation generalises without any new algebraic obstruction
aside from the opacity of heavily indexed notation:
the reduction formula for $I_n$ to $I_{n-1}$ with a 
choice of basis corresponding to 
a given choice of complementary $L+2$-gon is
\begin{align}
	I_{n}={1\over 2}\sum_{j\in\mathcal{E}_{L+2}}\xi^{[k_1,...,k_{L-2}]}_jI_{n-1}^{(j)} \; ,
	\quad {\rm for} \; n\geq L+2 \; ,
\end{align}
where $\lbrace k_1,...,k_{n}\rbrace = \lbrace 1,...,n\rbrace \slash \lbrace i_1,...,i_{n-L-2}\rbrace $,
the $i_n$s being the pinched propagators from the $n$-gon to form the 
$(L+2)$-gon\footnote{This is a more compact labelling of each 
$(L+1)$-gon in the large-$n$ case,
but less in the small-$n$ case as, 
for example, $I_5^{(1)} \equiv I_5^{[1,2,3,4,5]} $ 
for a hexagon descendant}.
For example, when $L=4$, the choice of reduction basis of $I_n$ corresponds to choosing the hexagon
depicted in figure~\ref{fig:hex}.

The general formula for the reduction to $(L+1)$-gons is
\begin{align}
	I_n &= {1\over 2^{n-L-1}}\sum_{i_1,...,i_{n-L-1}=1}^{n}\left[\prod_{m}^{n-L-1}
	\xi_{i_m}^{\left[\mathcal{P}, i_m\right]}\right]
	I_{L+1}^{\left[\mathcal{P} \right]} \; ,
	\quad {\rm for} \; n\geq L+2 \; ,
	\notag \\
	\mathcal{P}_{i} &= \left\lbrace  k_1,k_2,...,k_{L+1}\right\rbrace = \lbrace 1,...,n\rbrace\slash 
	\left\lbrace i_1,...,i_{n-L-1}\right\rbrace\quad .
	\label{eq:genred}
\end{align}

For the $L=4$ equal-internal-mass case\footnote{
It is a remarkable feature of $\tr_5$ that the expression
in equation~\ref{eq:hexi} is so broadly applicable.}:
\begin{equation}
	\xi^{[k_1,k_2,k_3,k_4,k_5,k_6]}_{k_j} = (-1)^{j}{2\tr_5 \left(q_{k_{j+1}k_{j+2}} 
	q_{k_{j+2}k_{j+3}}q_{k_{j+3}k_{j+4}}q_{k_{j+4}k_{j+5}}\right) 
	\over 
	\tr_5 \left(q_{k_1k_2} q_{k_2k_3}q_{k_3k_4}q_{k_4k_5}q_{k_5k_6}q_{k_6k_1}\right) }
	\; ,
	\label{eq:hexi}
\end{equation}
with the $js$ defined cyclically over $\lbrace 1, ..., 6\rbrace $.

\subsection{Feynman parameters in the numerator}
It is sometimes necessary 
to reduce $n$-gons with Feynman parameters
in the numerator.
In~\cite{Bern:1993kr},
a formula is derived for the reduction of an $n$-point
integral with a pair of Feynman parameters in the numerator.
We repeat the derivation here for completeness. 

We begin with a method analagous to used in section~\ref{sec:rev}
\begin{equation}
	J_{n;i}\equiv \Gamma (n-1-D/2)
	\int_0^1\cdots\int_0^{1-a_1-a_2-...\hat{a}_{i}...-a_{n-1}}
	da_n
	{d\over da_n} \left[ {a_k\over (\mathbf{a}^{{\rm T}}\Ss \mathbf{a})^{n+1-D/2}}\right] \biggr|_{a_i=1-a_1...-a_n} \;
	;
\end{equation}
the $\hat a_i$ signifies the omission of $a_i$;
applying the derivative gives
\begin{equation}
	J_{n;i,k} = \left(\delta_{nk}-\delta_{ik}\right)I_n^{D+2}
	-2\sum_{j=1}^n\left(\Ss_{nj}-\Ss_{ij}\right)I_n[a_ja_k]\quad .
	\label{eq:diffo}
\end{equation}
Alternatively, integration gives
\begin{equation}
	J_{n;i} = I_{n-1}^{(i)}\left[a_k\right] - I_{n-1}^{(n)}\left[a_k\right] \; ,
	\label{eq:partso}
\end{equation}
where $I^{(k)}[a_k]=0$.

Now combining \ref{eq:diffo} and \ref{eq:partso} allows us to 
define 
\begin{equation}
\Phi_k \equiv \delta_{ik}I_n^{D+2}-2\sum_{j=1}\Ss_{ij}I_n[a_ja_k]
	+I_{n-1}^{(i)}\left[a_k\right]
\end{equation}
which is independent of the choice of index $i$ on the right-hand 
side.

We assume $\Upsilon_n\neq 0$ and $\Delta_n\neq 0$; using the trick
\begin{equation}
	\sum_j	\Ss^{-1}_{ij} \Phi_k = \Ss^{-1}_{ik} I_n^{D+2} - 
	2I_n[a_ia_k]+\Ss_{ij}^{-1}I_{n-1}^{(j)}[a_k]
	\; ,
	\label{eq:pamo}
\end{equation}
where, also
\begin{equation}
	\sum_j	\Ss^{-1}_{ij} \Phi_k = c_i\Phi_k \; ,
\end{equation}
and using the fact that $\sum_j\Ss_{lj}c_j = 1$, then
\begin{equation}
	c_0 \Phi_k = c_0 I_n^{D+2}-2I_n[a_k]+c_j I_{n-1}^{(j)}\left[a_k\right] \quad .
\end{equation}
Combining with equations~\ref{eq:pamo} and~\ref{eq:red} gives
\begin{align}
	I_n[a_ia_k] = &{1\over 2} \left[ \Ss_{ik}^{\mi 1} + (n-2-D){c_ic_k\over c_0} \right] I_n^{D+2}
	\notag \\
	&+{1\over 4}(n-2-D)\left[ c_k \Ss_{ij}^{\mi 1} + c_i \Ss_{kj}^{\mi 1}- {c_ic_kc_j\over c_0} \right]
	I_{n-1}^{(j);D+2}
	+&{1\over 4}\Ss_{ij}^{\mi 1}\Ss_{kl}^{\mi 1}I_{n-2}^{(j,l)} \; ,
	\label{eq:2pam}
\end{align}
where we implicitly sum over $j$ consistent with summation convention.
Note that equation~\ref{eq:2pam} applies for all cases where $n\leq L+1$; 
a good consistency check can be carried out by 
confirming that summing equation~\ref{eq:2pam},
$\sum_{j=1}^nI_n[a_ja_i]=I_n[a_i]$, reduces to 
equation~\ref{eq:onepam}.

For $S_{ik}\neq 0$ we identify
\begin{align}
	I_n[a_ia_k]&= -{1\over 2}{\partial \over \partial S_{ik}} I^{D+2}_n[1] \; ,
	\label{eq:pam2der}
\end{align}
while for a $\Ss_{ik}=0$, these can be related to a non-zero entry through 
$\Phi_k$ equivalence.

The $n=L+2$ ($\Delta_n=c_0=0$) case can also be derived, 
\begin{align}
	I_{L+2}[a_ia_k] &=-{1\over 2} 
	{\partial \over \partial \Ss_{ik}}\left( I_{L+2}^{D+2}\right)
	={1\over 2}\left[ \sum_{j=1}^{L+2} \xi_j I^{(j)}_{L+1}[a_ia_k]
	-{1\over 2}{\sum_{j=1}^{L+2}{\partial \xi_j\over \partial \Ss_{ik}}I^{(j)}_{L+1}[1]} \right]\; ;
	\label{eq:2pamdeg}
\end{align}
and making the observation that
\begin{align}
	(\Ss \mathbf \xi)_l &= 1
\\ 
	\Rightarrow \left( {\partial \Ss \mathbf \xi \over \partial 
	\Ss_{ik}}\right)_l &= 0 
	\notag \\
	\Rightarrow 
	{\partial \xi_l\over \partial\Ss_{ik}} &=
	-\left( \Ss^{\mi 1}_{il}\xi_k + \Ss^{\mi 1}_{kl}\xi_i\right)\; 
	\label{eq:partder}
\end{align}
gives the generic formula for a monomial of Feynman parameters in terms 
of lower-point amplitude.

The reduction of any $n$-point integral for $n> L+2$ follows by replacing the reduction in
equation~\ref{eq:2pamdeg} by the relevant choice of reduction formulae; 
the matrix $S$ used in the simplification of the derivative in equation~\ref{eq:partder}
needs to be replaced by the appropriate $L+2$-gon $S_{ij}$ matrix for each $\xi$,
as described in the previous section.

Note that combining equation~\ref{eq:2pam} with 
equation~\ref{eq:pam2der}
generates the
partial differential equations for $I_n^{D+2}$ in terms of the
variables $\Ss_{ik}$.

We also note here that 
the analysis generalises to terms with a greater number
of Feynman parameters in the numerator, $N$, by generalising
$\Phi_k\rightarrow \Phi_{\mathbf{k}}\equiv \Phi_{k_1k_2...k_N}$
\begin{align}
	\Phi_{\mathbf{k}}  
	\equiv \sum_{j=1}^N\delta_{ik_l}I_n^{D+2}
	\left[a_{k_1}a_{k_2}\cdots \hat{a}_{k_l}\cdots a_{k_N}
	\right]
	-2\Ss_{ij}I_n\left[a_j\prod_{l=1}^N a_{k_l}\right]+
	I_{n-1}^{(i)}
	\left[\prod_{l=1}^N a_{k_l}\right] \quad .
\end{align}
We leave such derivations to the reader.
\newpage
\section{Maximal cuts}
\label{sec:cuts}
\begin{figure}[h]
\centerline{
    \begin{picture}(120,150)(-0,-50)    
     \Line( 0, 0)( 0,60)
     \Line( 0,60)(60,60)
     \Line(60,60)(60, 0)
     \Line(60, 0)( 0, 0)
     \Line( 0, 0)(-15,-15)
     \Line( 0,60)(-15,75)
     \Line(60,60)(60,78)
     \Line(60,60)(78,60)
     \Line(60, 0)(75,-15)
   \CCirc(60,60){4}{Black}{Black} 
     \Text(-20,-20)[c]{$1$}   
     \Text(-20,80)[c]{$2$}   
     \Text(63,82)[l]{$3$}  
     \Text(82,63)[l]{$4$}   
      \Text(80,-20)[c]{$5$}   
\SetWidth{2}
	    \SetColor{Red}
     \DashLine(30,-10)(30,10){2}
          \DashLine(30,50)(30,70){2}
               \DashLine(-10,30)(10,30){2}     
               \DashLine(50,30)(70,30){2}
    \end{picture} 
    }
	\caption[The quadruple cut]{The quadruple cut can be solved and the solutions plugged into the pinched propagator
	to give $c_4$. }
    \label{fig:box}
\end{figure}
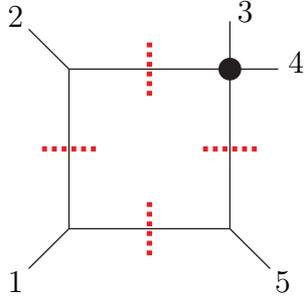

The Petersson formula~\ref{eq:peter} implies there is another way of finding $\mathbf{c}$
by solving the on-shell conditions of the one-mass boxes of the kind depicted in figure~\ref{fig:box}.

Without loss of generality, we consider $c_4$; 
the on-shell conditions read
\begin{align}
	\ell^2 &=0 \; ,
	\\
	(\ell-p_1)^2 &=0 \; ,
	\\
	(\ell-p_1-p_2)^2 &=0 \; ,
	\\
	(\ell+p_5)^2 &=0 \; ;
	\label{eq:onsh}
\end{align}
their two solutions lie in complex momentum space,
expressed in the spinor-helicity formalism as
\begin{align}
	\ell_{a\dot b} = l^+_{a\dot b} &= {\spb 1.2\over \spb 5.2}\lambda_1\bar\lambda_5 \; ,
	\\
	l^-_{a\dot b} &= {\spa 1.2\over \spa 5.2}\lambda_5\bar\lambda_1 \quad .
\end{align}
Applying these solutions into the pinched propagator:
\begin{align}
	(l^+-q_3)^2 &= {[2|34|5]\over \spb 5.2	}\; ,
	\notag \\ 
	(l^--q_3)^2 &= {\la 2|34|5 \ra \over \spa 5.2	}\; .
\end{align}
so that, using the definition~\ref{eq:peter}, 
\begin{align}
	c_4 &= {\spb 5.2\over [2|34|5]} +{\spa 5.2 \over \la 2|34|5 \ra} 
	\notag \\
	& = {\tr(2345)\over s_{23}s_{34}s_{45}} \;  
\end{align}
which matches equation~\ref{eq:cpent}, 
showing consistency between the ``Feynman-parametric", 
$S_{ij}$-matrix-based Melrose 
picture~\cite{Melrose1965} and the 
unparametrised 
cut-constraint based Petersson analysis~\cite{Petersson1965}.
\subsection{The $D$-dimensional pole}
Here we look at the singular behaviour missed by the four-dimensional cut constraints.
\begin{figure}[ht]
\centerline{
	\begin{picture}(100,100)(-50,-50)    
	\Line(-30,-40)(-40,-50)
	\Text(-43,-50)[r]{$1$}   
	\Line(40,-50)(30,-40)
	\Text(43,-50)[l]{$5$}   
	\Line(-61,7)(-45,5)
	\Text(-64,7)[r]{$2$}   
	\Line(61,7)(45,5)
	\Text(64,7)[l]{$4$}   
	\Line(0,57)(0,40)
	\Text(3,55)[l]{$3$}   
\SetWidth{2}
	\Line(-30,-40)(30,-40)
	\Line(45,5)(30,-40)
	\Line(-30,-40)(-45,5)
	\Line(0,40)(-45,5)
	\Line(45,5)(0,40)
\SetWidth{1}
	    \SetColor{Blue}
     \Line(0,-50)(0,-30)
     \Line(-50,-20)(-25,-11)
     \Line(50,-20)(25,-11)
     \Line(-30,34)(-11,14)
     \Line(30,34)(11,14)
    \end{picture} 
    }
	\caption[The penta cut]{The pentagon $D$-dimensional pole corresponds to the unique solution $l_0$ to the
	cut constraints~\ref{eq:5cut}}
    \label{fig:pent}
\end{figure}
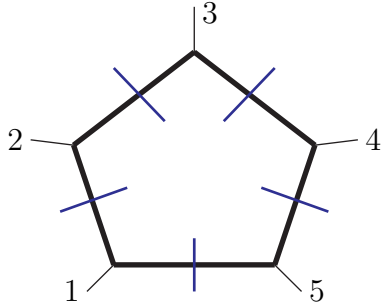    
Again specifying to the $L-4$ case, the $4-2\epsilon$ pentagon dimension-shift identity~\ref{eq:red} 
is
\begin{align}
	I_5= {1\over 2}\left[\sum_{j=1}^nc_jI^{(j)}_{4}
	+\epsilon c_0I^{6-2\epsilon}_5\right]\quad  .
		\label{eq:red5}
\end{align}
This equation has interesting properties when one considers the maximal cut of
the pentagon.
Cut conditions for the equal internal mass pentagon depicted in figure~\ref{fig:pent} are 
\begin{align}
	\ell^2 - M^2 = \ell^{[4]\; 2} + \ell^{[-2\epsilon ]\; 2} - M^2= l^2-\mu^2 &= 0 \; ,
	\notag \\
	(l-p_1)^2 - \mu^2 &= 0 \; ,
	\notag \\
	(l-p_1-p_2)^2 - \mu^2 &= 0 \; ,
	\notag \\ 
	(l+p_5+p_4)^2 -\mu^2 &= 0 \; ,
	\notag \\
	(l+p_5)^2 - \mu^2 &= 0 \quad ;
	\label{eq:5cut}
\end{align}
where we have defined $\mu^2 = M^2 - \ell^{[-2\epsilon ]\; 2}$ .
The unique solution to the conditions~\ref{eq:5cut} is 
\begin{align}
	l_0^\mu &=  {\tr_5\left(\gamma^\mu 12345\right) \over 2 \tr_5 (2345)}
	\label{eq:lpole}
	\\
	\mu_0^2 &= l^2 =  {16\Upsilon \over \Delta_5} = {1\over c_0 }\quad .
	\label{eq:mupole}
\end{align}
where in the spinor helicity formalism, if e.g. the external momenta 5 and 1 are massless ($m_1^2=m_2^2=0$), 
\begin{equation}
	{\tr_5\left(\sigma^{\dot{a}b} 12345\right)} = \lambda_1^{\dot{a}}[1|234|5\ra\bar{\lambda}_5^b - 
							\lambda_5^{\dot{a}}[5|432|1\ra\bar{\lambda}_1^b
\end{equation}
which can comfortably be generalised to the massive case over 
a given basis using the various 
schemes available~\cite{Dittmaier:1998nn,Schwinn2005,Cheung2009,Arkani-Hamed:2017jhn}.

Considering the integral explicitly and
re-introducing the normalisation of equation~\ref{eq:masint}
\begin{align}
	&(2\pi)^{4-2\epsilon}\mathcal{I}_5 =
	\int{}
	{d^{4-2\epsilon}\ell\over (\ell^2-M^2)((\ell-q_1)^2-M^2)((\ell-q_2)^2-M^2)((\ell-q_3)^2-M^2)((\ell-q_{4})^2-M^2)} \; ,
\end{align}
we follow the standard procedure of separating the integral measure:
\begin{align}
	d^{4-2\epsilon}\ell = -{1\over 2}\Omega_{S_{-2\epsilon}}{d^4l d\mu^2\over (\mu^2)^{1+\epsilon}}
\end{align}
where $\Omega_{S_{\mi 2\epsilon}}$ is the unit surface area
of the sphere, $S_{\mi 2\epsilon}$, as is normally 
extracted from the loop integration step.
Upon shifting the $\mu^2$ variable such that it matches the definition
from the conditions~\ref{eq:5cut},
this gives the integral
\begin{align}
	=&-{1\over 2}\Omega_{S_{-2\epsilon}}\int_{M^2}^\infty {d\mu^2\over (\mu^2-M^2)^{1+\epsilon}}
	\times
	\notag \\ 
	&\hspace{0.5in}\int{{d^4l}\over 
	(l^2-\mu^2)((l-q_1)^2-\mu^2)((l-q_2)^2-\mu^2)((l-q_3)^2-\mu^2)((l-q_{4})^2-\mu^2)} \quad .
\end{align}
The residues of the integrand can be captured 
changing the $\mu^2$ integration region to 
a large enough contour on the complex plane,
and taking the limit $\epsilon\rightarrow 0$; 
we can see that where the ``four-dimensional" residue is simply
manifestly captured by the pole at
\begin{align}
	\mu^2 = M^2
\end{align}
and need only be combined with four other conditions 
(for example the box cut conditions~\ref{eq:onsh}). 

The $D$-dimensional pole, on the other hand,  
is buried in the solution to the on-shell conditions~\ref{eq:5cut};
it lies at
\begin{align}
	\mu^2 = {1\over c_0} \; .
\end{align}
It is not a coincidence that this is the the inverse of 
the coefficient  
of the shifted integral in equation~\ref{eq:red}.
as indicated by Schnetz for the Petersson formula in the integer dimension case~\cite{Schnetz2010}, 
one can view 
the cut version of equation~\ref{eq:red5} as being analagous to a sum over residues;
it can be seen as a ``freezing out" of propagators into rational coefficients. 

This analysis can be carried forward to the $n$-gon reduction formula~\ref{eq:genred};
this is straightforward as we observe that in the hexagon case
\begin{equation}
	I_6 = {1\over 2}\sum_{j=1}^6 \xi_j I_5^{(j)} \; ,
\end{equation}
taking the solution~\ref{eq:lpole} for e.g. $I_5^{(6)}$ gives
\begin{align}
	(\ell+p_6) = 2 l_0\cdot p_6 + p_6^2 = {\tr_5\left(612345\right) \over  \tr_5 (1234)} =
{2\over \xi_6}  \quad .
\end{align}

Consistent with the Petersson formula for a hexagon, 
the coefficient of the descendant box solutions satisfy the equations
\begin{equation}
	{1\over 2}\left(\xi_ic^{(i)}_j+\xi_jc^{(j)}_i \right) = {1\over \pr_i(l^+)\pr_j(l^+)}+
					{1\over \pr_i(l^-)\pr_j(l^-)}\; , 
					\label{eq:pet}
\end{equation}
where $\pr_i(l^{\pm})$ is the inverse ``frozen out" $i$th propagator with the null 
cut solutions of the other propagators $l^\pm$ inserted. 
Note that the Petersson form on the right-hand side of equation~\ref{eq:pet}
does not factor in such a simple manner, but 
is averaged over the all positive and all negative products.
In this sense the unique solution to the $D$-dimensional constraint is simpler
as it avoids the need to average:
\begin{equation}
	\prod_{m}^{n-L-1}
	\xi_{i_m}^{\left[\mathcal{P}, i_m\right]}
	=\prod_{m}^{n-L-1}{2^{n-L-2}\over  \pr (\ell_{I_m})}
	\quad .
\end{equation}

\section{Concluding remarks}
The formulae presented in this paper for the reduction of integrals should greatly simplify high-multiplicity amplitude computations.
As well as being useful for reducing high-multiplicity integrals, the reduction formulae~\ref{eq:genred}
also presents a very simple interpretation in terms of ``freezing out" of propagators.

This integration with unitarity cuts leaves open the prospect of computing all-epsilon, 
all-multiplicity closed forms
for scattering amplitudes, which as well as being a step forwards methodically,
could also further more general analytic understanding of scattering amplitudes.
The remarkable simplicity of the solution to the pentagon equal-mass conditions~\ref{eq:5cut}
opens up the prospect of a ``generalised unitarity" approach to $D$-dimensional unitarity, 
avoiding the need for integrand reconstruction.

One could pose the question as to whether the ``freezing out" can be extended to simplify multi-loop relations 
where reductions
and relations between integrals are more involved.

We leave these questions to future work.

\section*{Acknowledgements}

I would like to thank Ruth Britto, Riccardo Gonzo, Martijn Hidding, Andrea Orta and Johann Usovitsch for
countless useful conversations during the gestation of this work.
Particular thanks to Anne Spiering for helpful comments on 
a draft of this manuscript.
Thank you also to Ruth Britto for looking over the paper.

I would also like to thank the Galileo Galilei Institute (GGI) for hosting me for the ``Amplitudes in the LHC era"  
workshop where some of the initial parts of this work developed.

Diagrams were drawn using Axodraw.

This work was supported by the ERC ``CutLoops" consolidator grant number 647356.

\appendix
\section{Explicit derivations}
\subsection{$n=L+3$ reduction}
\label{app:der}
Beginning with equation~\ref{eq:startder}, we have
\begin{align}
	-I_{L+2}^{(i)}[1]&= \left(L+1-D\right)
		I^{D+2}_{L+3}[1]
		- 2\sum_{j=1}^n\Ss_{ij}I_{L+3}[a_j]\; ,
\end{align}
and then multiplying by $\xi^{(k)}_i$ and summing
\begin{align}
	{1\over 2}\sum_{i\in\mathcal{E}_{L+2}}\xi^{(k)}_iI_{L+2}^{(i)}[1]&= 
	\sum_{i\in\mathcal{E}_{L+2}}\xi_i^{(k)}\sum_{j=1}^n\Ss_{ij}I_{L+3}[a_j]\; 
		\notag \\
	&= \sum_{i\in\mathcal{E}_{L+2}}\xi^{(k)}_i\sum_{j\in\mathcal{E}_{L+2}}^n\Ss_{ij}I_{L+3}[a_j]\;
	+ \sum_{i\in\mathcal{E}_{L+2}}\xi^{(k)}_i\Ss_{ik}I_{L+3}[a_k]\; ,
\end{align}
which, using the identity
\begin{equation}
	\sum_{i\in \mathcal{E}} \Ss_{ij} \xi_{i}^{(k)} = 1 \; ,
\end{equation}
and $\sum_{j=1}^n a_j =1$ gives
\begin{align}
	{1\over 2}\sum_{j\in\mathcal{E}_{L+2}}\xi^{(k)}_iI_{L+2}^{(i)}[1]&= 
	 I_{L+3}[1]
	+ \left(\sum_{i\in\mathcal{E}_{L+2}}\xi^{(k)}_i\Ss_{ik}-1\right)I_{L+3}[a_k]\quad .
\end{align}
Recalling equation~\ref{eq:cno} (and defining $\xi^{(k)}_k=0$)
\begin{align}
	\sum_{i=1}^{L+3}\xi^{(k)}_i&=2^{L+1} 
	{\Delta^{(k)}_{L+2}\over
	\Upsilon^{(k)}_{L+2}}= 0 \; ,
\end{align}
and, as $\Upsilon_{L+3}=0$, 
\begin{align}
	\Ss_{ik}=\Ss_{i(k+1)}+
	 {\rm column \; operations } \; ,
\end{align}
 then we deduce that
\begin{align}
	\sum_{i\in\mathcal{E}_{L+2}}\xi^{(k)}_i\Ss_{ik}-1 = 0
	\quad .
\end{align}
Thus
\begin{align}
	 I_{L+3}={1\over 2}\sum_{j\in\mathcal{E}_{L+2}}\xi^{(k)}_iI_{L+2}^{(i)} \; , 
\end{align}
which is equation~\ref{eq:endder}.

The reduction formula~\ref{eq:lred} can now be applied to each 
$(L+2)$-gon to give
\begin{align}
	I_{L+3}={1\over 4}\sum_{j,k=1}^n\xi^{(j)}_k\xi^{(k)}_jI_{L+1}^{(j,k)} \; ,
\end{align}
which is equation~\ref{eq:hept}.

\bibliography{biblo}{}

\begin{thebibliography}{10}
\providecommand{\href}[2]{#2}
\providecommand{\arxivref}[2]{\href{http://arxiv.org/abs/#1}{#2}}
\providecommand{\doiref}[2]{\href{http://dx.doi.org/#1}{#2}}
\providecommand{\nbbstauthor}[1]{#1}
\providecommand{\nbbstjournal}[1]{\textsf{#1}}
\providecommand{\nbbsttitle}[1]{\textit{#1}}
\providecommand{\nbbsturl}[1]{\texttt{#1}}
\providecommand{\nbbsteprint}[1]{\texttt{#1}}
\providecommand{\nbbststyle}{\raggedright\small\parskip0pt}
\nbbststyle

\bibitem{Badger:2015lda}
\nbbstauthor{S.~Badger, G.~Mogull, A.~Ochirov and D.~O'Connell},
\nbbsttitle{``{A Complete Two-Loop, Five-Gluon Helicity Amplitude in Yang-Mills
  Theory}''},
\nbbstjournal{\doiref{10.1007/JHEP10(2015)064}{JHEP~1510,~064~(2015)}},
\nbbsteprint{\arxivref{1507.08797}{arxiv:1507.08797}}.

\bibitem{Badger2018}
\nbbstauthor{S.~Badger, C.~Brønnum-Hansen, H.~B.~Hartanto and T.~Peraro},
\nbbsttitle{``{Analytic helicity amplitudes for two-loop five-gluon scattering:
  the single-minus case}''},
\nbbsteprint{\arxivref{1811.11699}{arxiv:1811.11699}}.

\bibitem{Abreu:2019odu}
\nbbstauthor{S.~Abreu, J.~Dormans, F.~Febres~Cordero, H.~Ita, B.~Page and
  V.~Sotnikov},
\nbbsttitle{``{Analytic Form of the Planar Two-Loop Five-Parton Scattering
  Amplitudes in QCD}''},
\nbbstjournal{\doiref{10.1007/JHEP05(2019)084}{JHEP~1905,~084~(2019)}},
\nbbsteprint{\arxivref{1904.00945}{arxiv:1904.00945}}.

\bibitem{Badger:2019djh}
\nbbstauthor{S.~Badger, D.~Chicherin, T.~Gehrmann, G.~Heinrich, J.~M.~Henn,
  T.~Peraro, P.~Wasser, Y.~Zhang and S.~Zoia},
\nbbsttitle{``{Analytic form of the full two-loop five-gluon all-plus helicity
  amplitude}''},
\nbbstjournal{\doiref{10.1103/PhysRevLett.123.071601}{Phys.~Rev.~Lett.~123,~071601~(2019)}},
\nbbsteprint{\arxivref{1905.03733}{arxiv:1905.03733}}.

\bibitem{Dunbar2020a}
\nbbstauthor{D.~C.~Dunbar, J.~H.~Godwin, W.~B.~Perkins and J.~M.~W.~Strong},
\nbbsttitle{``{Color Dressed Unitarity and Recursion for Yang-Mills Two-Loop
  All-Plus Amplitudes}''},
\nbbstjournal{\doiref{10.1103/PhysRevD.101.016009}{Phys.~Rev.~D101,~016009~(2020)}},
\nbbsteprint{\arxivref{1911.06547}{arxiv:1911.06547}}.

\bibitem{Caron-Huot:2019bsq}
\nbbstauthor{S.~Caron-Huot, L.~J.~Dixon, F.~Dulat, M.~Von~Hippel, A.~J.~McLeod
  and G.~Papathanasiou},
\nbbsttitle{``{The Cosmic Galois Group and Extended Steinmann Relations for
  Planar $\mathcal{N} = 4$ SYM Amplitudes}''},
\nbbstjournal{\doiref{10.1007/JHEP09(2019)061}{JHEP~1909,~061~(2019)}},
\nbbsteprint{\arxivref{1906.07116}{arxiv:1906.07116}}.

\bibitem{Bern:1994cg}
\nbbstauthor{Z.~Bern, L.~J.~Dixon, D.~C.~Dunbar and D.~A.~Kosower},
\nbbsttitle{``{Fusing gauge theory tree amplitudes into loop amplitudes}''},
\nbbstjournal{\doiref{10.1016/0550-3213(94)00488-Z}{Nucl.~Phys.~B435,~59~(1995)}},
\nbbsteprint{\arxivref{hep-ph/9409265}{hep-ph/9409265}}.

\bibitem{Mahlon:1993si}
\nbbstauthor{G.~Mahlon},
\nbbsttitle{``{Multi - gluon helicity amplitudes involving a quark loop}''},
\nbbstjournal{\doiref{10.1103/PhysRevD.49.4438}{Phys.~Rev.~D49,~4438~(1994)}},
\nbbsteprint{\arxivref{hep-ph/9312276}{hep-ph/9312276}}.

\bibitem{Mahlon:1994dc}
\nbbstauthor{G.~Mahlon},
\nbbsttitle{``{Use of recursion relations to compute one loop helicity
  amplitudes}''},
\nbbsteprint{\arxivref{hep-ph/9412350}{hep-ph/9412350}},
in: \nbbsttitle{``{Beyond the standard model 4. Proceedings, 4th International
  Conference on High-Energy Physics, Tahoe City, USA, December 13-18, 1994}''},
pp.~475-478.

\bibitem{Dunbar2020}
\nbbstauthor{D.~C.~Dunbar, W.~B.~Perkins and J.~M.~W.~Strong},
\nbbsttitle{``{An $n$-point QCD two-loop amplitude}''},
\nbbsteprint{\arxivref{2001.11347}{arxiv:2001.11347}}.

\bibitem{Anastasiou2003}
\nbbstauthor{C.~Anastasiou, Z.~Bern, L.~J.~Dixon and D.~A.~Kosower},
\nbbsttitle{``{Planar amplitudes in maximally supersymmetric Yang-Mills
  theory}''},
\nbbstjournal{\doiref{10.1103/PhysRevLett.91.251602}{Phys.~Rev.~Lett.~91,~251602~(2003)}},
\nbbsteprint{\arxivref{hep-th/0309040}{hep-th/0309040}}.

\bibitem{Bern2005}
\nbbstauthor{Z.~Bern, L.~J.~Dixon and V.~A.~Smirnov},
\nbbsttitle{``{Iteration of planar amplitudes in maximally supersymmetric
  Yang-Mills theory at three loops and beyond}''},
\nbbstjournal{\doiref{10.1103/PhysRevD.72.085001}{Phys.~Rev.~D72,~085001~(2005)}},
\nbbsteprint{\arxivref{hep-th/0505205}{hep-th/0505205}}.

\bibitem{Gehrmann:2015bfy}
\nbbstauthor{T.~Gehrmann, J.~M.~Henn and N.~A.~Lo~Presti},
\nbbsttitle{``{Analytic form of the two-loop planar five-gluon
  all-plus-helicity amplitude in QCD}''},
\nbbstjournal{\doiref{10.1103/PhysRevLett.116.189903,
  10.1103/PhysRevLett.116.062001}{Phys.~Rev.~Lett.~116,~062001~(2016)}},
\nbbsteprint{\arxivref{1511.05409}{arxiv:1511.05409}},
[Erratum: Phys. Rev. Lett.116,no.18,189903(2016)].

\bibitem{Dunbar:2016aux}
\nbbstauthor{D.~C.~Dunbar and W.~B.~Perkins},
\nbbsttitle{``{Two-loop five-point all plus helicity Yang-Mills amplitude}''},
\nbbstjournal{\doiref{10.1103/PhysRevD.93.085029}{Phys.~Rev.~D93,~085029~(2016)}},
\nbbsteprint{\arxivref{1603.07514}{arxiv:1603.07514}}.

\bibitem{Dunbar:2017nfy}
\nbbstauthor{D.~C.~Dunbar, J.~H.~Godwin, G.~R.~Jehu and W.~B.~Perkins},
\nbbsttitle{``{Analytic all-plus-helicity gluon amplitudes in QCD}''},
\nbbstjournal{\doiref{10.1103/PhysRevD.96.116013}{Phys.~Rev.~D96,~116013~(2017)}},
\nbbsteprint{\arxivref{1710.10071}{arxiv:1710.10071}}.

\bibitem{Henn:2019mvc}
\nbbstauthor{J.~Henn, B.~Power and S.~Zoia},
\nbbsttitle{``{Conformal Invariance of the One-Loop All-Plus Helicity
  Scattering Amplitudes}''},
\nbbstjournal{\doiref{10.1007/JHEP02(2020)019}{JHEP~2002,~019~(2020)}},
\nbbsteprint{\arxivref{1911.12142}{arxiv:1911.12142}}.

\bibitem{Catani:1998bh}
\nbbstauthor{S.~Catani},
\nbbsttitle{``{The Singular behavior of QCD amplitudes at two loop order}''},
\nbbstjournal{\doiref{10.1016/S0370-2693(98)00332-3}{Phys.~Lett.~B427,~161~(1998)}},
\nbbsteprint{\arxivref{hep-ph/9802439}{hep-ph/9802439}}.

\bibitem{Bern:1996ja}
\nbbstauthor{Z.~Bern, L.~J.~Dixon, D.~C.~Dunbar and D.~A.~Kosower},
\nbbsttitle{``{One loop selfdual and N=4 superYang-Mills}''},
\nbbstjournal{\doiref{10.1016/S0370-2693(96)01676-0}{Phys.~Lett.~B394,~105~(1997)}},
\nbbsteprint{\arxivref{hep-th/9611127}{hep-th/9611127}}.

\bibitem{Britto:2020crg}
\nbbstauthor{R.~Britto, G.~R.~Jehu and A.~Orta},
\nbbsttitle{``{The dimension-shift conjecture for one-loop amplitudes}''},
\nbbsteprint{\arxivref{2011.13821}{arxiv:2011.13821}}.

\bibitem{Brown1961}
\nbbstauthor{L.~M.~Brown},
\nbbsttitle{``{Analytic properties of npoint loops in perturbation theory}''},
\nbbstjournal{\doiref{10.1007/BF02829004}{Nuovo~Cim.~22,~178~(1961)}}.

\bibitem{Halpern1963}
\nbbstauthor{F.~R.~Halpern},
\nbbsttitle{``{Reduction formula for the five-point function}''},
\nbbstjournal{\doiref{10.1103/PhysRevLett.10.310}{Phys.~Rev.~Lett.~10,~310~(1963)}}.

\bibitem{Melrose1965}
\nbbstauthor{D.~B.~Melrose},
\nbbsttitle{``{Reduction of Feynman diagrams}''},
\nbbstjournal{\doiref{10.1007/BF02832919}{Nuovo~Cim.~40,~181~(1965)}}.

\bibitem{Petersson1965}
\nbbstauthor{B.~Petersson},
\nbbsttitle{``{Reduction of a one-loop Feynman diagram with n vertices in
  m-dimensional Lorentz space}''},
\nbbstjournal{\doiref{10.1063/1.1704747}{J.~Math.~Phys.~6,~1955~(1965)}}.

\bibitem{Neerven1984}
\nbbstauthor{W.~L.~van~Neerven and J.~A.~M.~Vermaseren},
\nbbsttitle{``{Large loop integrals}''},
\nbbstjournal{\doiref{10.1016/0370-2693(84)90237-5}{Phys.~Lett.~137B,~241~(1984)}}.

\bibitem{Bern:1992em}
\nbbstauthor{Z.~Bern, L.~J.~Dixon and D.~A.~Kosower},
\nbbsttitle{``{Dimensionally regulated one loop integrals}''},
\nbbstjournal{\doiref{10.1016/0370-2693(93)90469-X,
  10.1016/0370-2693(93)90400-C}{Phys.~Lett.~B302,~299~(1993)}},
\nbbsteprint{\arxivref{hep-ph/9212308}{hep-ph/9212308}},
[Erratum: Phys. Lett.B318,649(1993)].

\bibitem{Bern:1993kr}
\nbbstauthor{Z.~Bern, L.~J.~Dixon and D.~A.~Kosower},
\nbbsttitle{``{Dimensionally regulated pentagon integrals}''},
\nbbstjournal{\doiref{10.1016/0550-3213(94)90398-0}{Nucl.~Phys.~B412,~751~(1994)}},
\nbbsteprint{\arxivref{hep-ph/9306240}{hep-ph/9306240}}.

\bibitem{Tarasov1996d}
\nbbstauthor{O.~V.~Tarasov},
\nbbsttitle{``{Connection between Feynman integrals having different values of
  the space-time dimension}''},
\nbbstjournal{\doiref{10.1103/PhysRevD.54.6479}{Phys.~Rev.~D54,~6479~(1996)}},
\nbbsteprint{\arxivref{hep-th/9606018}{hep-th/9606018}}.

\bibitem{Lee2010a}
\nbbstauthor{R.~N.~Lee},
\nbbsttitle{``{Space-time dimensionality D as complex variable: Calculating
  loop integrals using dimensional recurrence relation and analytical
  properties with respect to D}''},
\nbbstjournal{\doiref{10.1016/j.nuclphysb.2009.12.025}{Nucl.~Phys.~B830,~474~(2010)}},
\nbbsteprint{\arxivref{0911.0252}{arxiv:0911.0252}}.

\bibitem{Lee2010b}
\nbbstauthor{R.~N.~Lee, A.~V.~Smirnov and V.~A.~Smirnov},
\nbbsttitle{``{Dimensional recurrence relations: an easy way to evaluate higher
  orders of expansion in $\epsilon$}''},
\nbbstjournal{\doiref{10.1016/j.nuclphysbps.2010.09.011}{Nucl.~Phys.~Proc.~Suppl.~205-206,~308~(2010)}},
\nbbsteprint{\arxivref{1005.0362}{arxiv:1005.0362}},
in: \nbbsttitle{``{Proceedings, 10th DESY Workshop on Elementary Particle
  Theory: Loops and Legs in Quantum Field Theory: Woerlitz, Germany, April
  25-30, 2010}''},
pp.~308-313.

\bibitem{Lee2012}
\nbbstauthor{R.~N.~Lee and V.~A.~Smirnov},
\nbbsttitle{``{The Dimensional Recurrence and Analyticity Method for
  Multicomponent Master Integrals: Using Unitarity Cuts to Construct
  Homogeneous Solutions}''},
\nbbstjournal{\doiref{10.1007/JHEP12(2012)104}{JHEP~1212,~104~(2012)}},
\nbbsteprint{\arxivref{1209.0339}{arxiv:1209.0339}}.

\bibitem{Bytev2014}
\nbbstauthor{V.~V.~Bytev, M.~{\relax Yu}.~Kalmykov and S.-O.~Moch},
\nbbsttitle{``{HYPERgeometric functions DIfferential REduction (HYPERDIRE):
  MATHEMATICA based packages for differential reduction of generalized
  hypergeometric functions: $F_D$ and $F_S$ Horn-type hypergeometric functions
  of three variables}''},
\nbbstjournal{\doiref{10.1016/j.cpc.2014.07.014}{Comput.~Phys.~Commun.~185,~3041~(2014)}},
\nbbsteprint{\arxivref{1312.5777}{arxiv:1312.5777}}.

\bibitem{Bern:1991aq}
\nbbstauthor{Z.~Bern and D.~A.~Kosower},
\nbbsttitle{``{The Computation of loop amplitudes in gauge theories}''},
\nbbstjournal{\doiref{10.1016/0550-3213(92)90134-W}{Nucl.~Phys.~B379,~451~(1992)}}.

\bibitem{Binoth2000}
\nbbstauthor{T.~Binoth, J.~P.~Guillet and G.~Heinrich},
\nbbsttitle{``{Reduction formalism for dimensionally regulated one loop N point
  integrals}''},
\nbbstjournal{\doiref{10.1016/S0550-3213(00)00040-7}{Nucl.~Phys.~B572,~361~(2000)}},
\nbbsteprint{\arxivref{hep-ph/9911342}{hep-ph/9911342}}.

\bibitem{Britto:2004nc}
\nbbstauthor{R.~Britto, F.~Cachazo and B.~Feng},
\nbbsttitle{``{Generalized unitarity and one-loop amplitudes in N=4
  super-Yang-Mills}''},
\nbbstjournal{\doiref{10.1016/j.nuclphysb.2005.07.014}{Nucl.~Phys.~B725,~275~(2005)}},
\nbbsteprint{\arxivref{hep-th/0412103}{hep-th/0412103}}.

\bibitem{Binoth2005}
\nbbstauthor{T.~Binoth, J.~P.~Guillet, G.~Heinrich, E.~Pilon and C.~Schubert},
\nbbsttitle{``{An Algebraic/numerical formalism for one-loop multi-leg
  amplitudes}''},
\nbbstjournal{\doiref{10.1088/1126-6708/2005/10/015}{JHEP~0510,~015~(2005)}},
\nbbsteprint{\arxivref{hep-ph/0504267}{hep-ph/0504267}}.

\bibitem{Dittmaier:1998nn}
\nbbstauthor{S.~Dittmaier},
\nbbsttitle{``Weyl-van-der-Waerden formalism for helicity amplitudes of massive
  particles''},
\nbbsteprint{\arxivref{hep-ph/9805445}{hep-ph/9805445}}.

\bibitem{Schwinn2005}
\nbbstauthor{C.~Schwinn and S.~Weinzierl},
\nbbsttitle{``{Scalar diagrammatic rules for Born amplitudes in QCD}''},
\nbbstjournal{\doiref{10.1088/1126-6708/2005/05/006}{JHEP~0505,~006~(2005)}},
\nbbsteprint{\arxivref{hep-th/0503015}{hep-th/0503015}}.

\bibitem{Cheung2009}
\nbbstauthor{C.~Cheung and D.~O'Connell},
\nbbsttitle{``{Amplitudes and Spinor-Helicity in Six Dimensions}''},
\nbbstjournal{\doiref{10.1088/1126-6708/2009/07/075}{JHEP~0907,~075~(2009)}},
\nbbsteprint{\arxivref{0902.0981}{arxiv:0902.0981}}.

\bibitem{Arkani-Hamed:2017jhn}
\nbbstauthor{N.~Arkani-Hamed, T.-C.~Huang and Y.-t.~Huang},
\nbbsttitle{``{Scattering Amplitudes For All Masses and Spins}''},
\nbbsteprint{\arxivref{1709.04891}{arxiv:1709.04891}}.

\bibitem{Schnetz2010}
\nbbstauthor{O.~Schnetz},
\nbbsttitle{``{The geometry of one-loop amplitudes}''},
\nbbsteprint{\arxivref{1010.5334}{arxiv:1010.5334}}.

\end{thebibliography}
\bibliographystyle{nb.bst}

\end{document}